\documentclass[a4paper,11pt]{article}
\usepackage{amsfonts}

\usepackage{graphicx}
\graphicspath{{Figures/}}
\usepackage{amsmath}
\usepackage{amssymb}
\usepackage{latexsym}
\usepackage[colorlinks]{hyperref}
\usepackage{color}
\usepackage{float}
\usepackage{cite}
\usepackage{makeidx}
\usepackage{colortbl}
\usepackage{csquotes}
\usepackage[squaren]{SIunits}

\usepackage[textfont={footnotesize,sf},labelfont={color=blue,bf,sf},labelsep=endash]{caption}
\usepackage[position=top,labelfont={color=blue,bf,sf}]{subfig}
\usepackage{bm}

\usepackage[title]{appendix}

\newcommand{\bra}{\begin{array}}
    \newcommand{\era}{\end{array}}
\newcommand{\beq}{\begin{equation}}
    \newcommand{\eeq}{\end{equation}}
\newcommand{\bqr}{\begin{eqnarray}}
    \newcommand{\eqr}{\end{eqnarray}}

\newcommand{\lb}{\label}

\begin{document}
    \begin{titlepage}
        \setcounter{page}{1}
        \renewcommand{\thefootnote}{\fnsymbol{footnote}}

        \begin{flushright}
        \end{flushright}

        \vspace{5mm}
        \begin{center}

            {\Large \bf {Gap-tunable of  
                    Tunneling Time in Graphene Magnetic Barrier}}

            \vspace{5mm}

            {\bf   Youssef Fattasse}$^{a}$, {\bf Miloud Mekkaoui}$^{a}$,  {\bf Ahmed Jellal\footnote{\sf a.jellal@ucd.ac.ma}}$^{a,b}$ and {\bf Abdelhadi Bahaoui}$^{a}$

            \vspace{5mm}

            {$^a$\em Laboratory of Theoretical Physics, Faculty of Sciences, Choua\"ib Doukkali University},\\
            {\em PO Box 20, 24000 El Jadida, Morocco}

            {$^{b}$\em Canadian Quantum  Research Center,
                204-3002 32 Ave Vernon, \\ BC V1T 2L7,  Canada}

            \vspace{3cm}

            \begin{abstract}

                We study the tunneling time of Dirac fermions in
                graphene  magnetic barrier through  an electrostatic potential and a mass term. This latter generates an energy gap in the spectrum and therefore affects the proprieties of tunneling of the system. For clarification,  we first start by
                deriving the  eigenspinors solutions of Dirac equation and second  connect  them to the incident, reflected and transmitted beam waves. This  connection allows us to obtain the corresponding phases shifts and consequently compute the group delay time in transmission and reflection.
                Our numerical
                results show that
                the group delay time
                depends strongly  on the energy gap  in the tunneling process
                through single barrier. Moreover, we find that the group
                approaches unity at some
                critical value of the energy gap and becomes independent to the strengths of  involved  physical parameters.
                
                \vspace{3cm}



            \end{abstract}
        \end{center}
    \end{titlepage}


    \section{ Introduction}

    In graphene sheets, the particles
    and density of the carriers can be controlled by tuning a gate
    bias voltage \cite{Novoselov1,Zhang1,Nilsson1}. In a gapless
    graphene electrical conduction cannot be switched off by using the
    control voltages \cite{ Katsnelson1} that is essential for the
    operation of conventional transistors.
    This situation can be solved by generating an energy gap in the electronic spectrum of the carriers in graphene. Such gap is a measure of the threshold voltage and the on-off
    ratio of the field effect transistors \cite{Lin1, Kedzierski1}.
    Experimentally, there are
different techniques  to open a gap in a graphene band structure \cite{3333} and its maximum value could be  260 meV because of the sublattice symmetry breaking \cite{2525}. We emphasis that its measurement varies from an experiment to another.  For instance,
an energy gap can be measured under the control
of
the structure of the interface between graphene and ruthenium \cite{2626}
or
by considering
 graphene grown epitaxially on a SiC substrate \cite{2525}. From theoretical point of view,
 alternative strategies have been proposed
 to generate an energy gap  in systems based on graphene. There is a vast  literature, but here we mention two developed strategies.
  Indeed, a  graphene sheet on top of a lattice-matched hexagonal boron nitride (h-BN) substrate leading to a gap of 53 meV
  \cite{2626}.
  Also
  a graphene subject to the potential from an external superlattice can develop
  a gap
  provided the superlattice potential has a  broken inversion symmetry
  \cite{Tiwari, Jellal}.


A basic quantum property is the tunnel effect that occurs when a particle passes through a potential barrier. Actually, it is known that under the increase of width and height of a barrier, the chance of a non-relativistic particle breaching it decreases exponentially. In contrast to the relativistic theory, which   predicted that no matter how tall or wide the barrier is, a relativistic  particle would tunnel through it with certainty. Such effect is  called Klein tunneling  that has been experimentally realized in graphene \cite{66,77,88,99}.
On the other hand,
the nanostructures based on graphene-magnetic barriers
have recently been the subject of several investigations focusing on the tunneling time by
treating time as a parameter rather than an observable in quantum mechanics \cite{144,155,166,177}.
As a result, there is no straightforward way to quantify tunneling time. There are at least three different notions of traversing time in the literature for particles with a given energy \cite{188}. To obtain the
group delay time (called also  phase time \cite{Hauge}),
 one first studies the evolution of the wave packets through the barrier, which requires the phase sensitivity of the tunneling amplitude to the incident  energy of particles.
In electron transport 
\cite{Buttiker,Gasparian,Buttiker2002}
 and photon tunneling 
 \cite{Steinberg,Spielmann}, the
 group delay time has been described as one of the most important and interesting quantities.
For instance, the group delay statistics are linked to dynamic admittance as well as other material microstructure properties \cite{Buttiker} and 
to density of states \cite{Gasparian,Buttiker2002}.

    We investigate the group delay time for graphene
    magnetic field in the presence of a barrier potential and mass term.
    As a result
    we solve the Dirac equation in three regions to
    determine the
    solutions of the energy spectrum in terms of the energy gap $ \Delta $.
     After matching the wave functions at  interfaces and using the density current,
     we calculate the transmission and reflection probabilities. By mapping the incident, reflected and transmitted beam waves as functions of the eigenspinors 
     we compute
     the group delay time. To give a better understanding of our results, we
     numerically investigate the basic features of the group and show that
     it can be controlled by tuning on  $ \Delta $.
     Moreover, we will discuss how the system
     parameters can affect the tunneling time in gapped graphene.
    In fact, we will study the main
    characteristics of this quantity 
    in terms
    of the physical parameter of our system.

    This paper is organized as follows. In section 2, we formulate our theoretical
    problem by writing the corresponding Hamiltonian
    and determine the eigenspinors and eigenvalues.
    In section 3,
    we use the boundary conditions together with  current density
    to compute  transmission and reflection probabilities that allow to derive the
    group delay time.
 %
    We numerically discuss our results by giving
    different illustrations under suitable choices of the physical parameters, in section 4.
    Finally, we close our work by summarizing the main obtained results.

    \section{ Theoretical model}

\begin{figure}[ht]
    \centering
    \subfloat[]{
        \centering
        \includegraphics[width=0.43\linewidth]{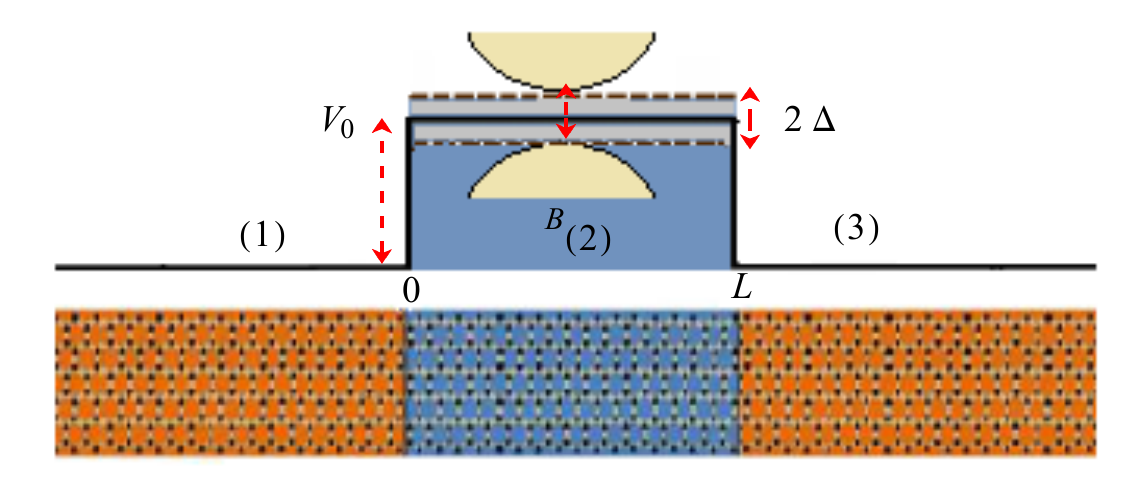}
        \label{db.3}\ \ \ \ \ \ \ \
    }\subfloat[]{
        \centering
        \includegraphics[width=0.27\linewidth]{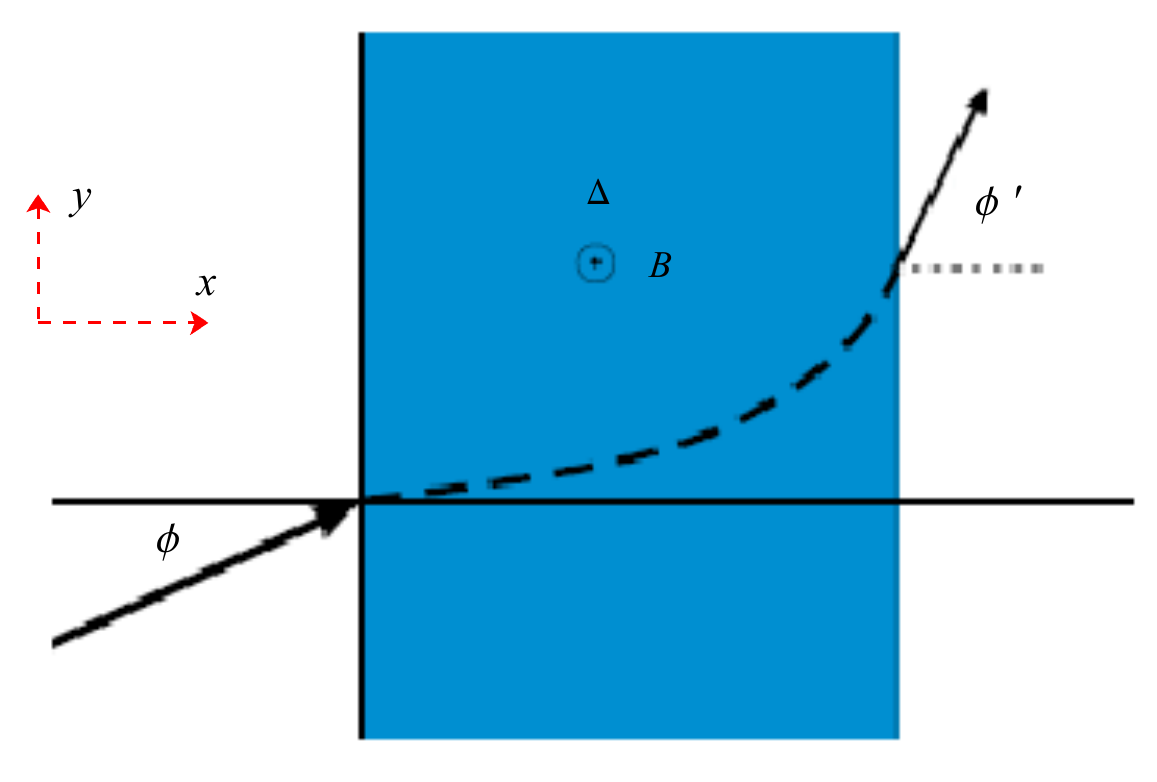}
        \label{db.4}}
    \caption{\sf{(color online) {\color{red}{(a)}}\color{black}{:} Schematic of the potential profile in  graphene
        magnetic barrier with mass term, forming  three regions denoted by $ j=1,2,3 $.
            {\color{red}{(b)}}\color{black}{:}   Trajectory for an electron of our system having the incident angle $\phi$ and the transmitted angle $\phi'$.
    }}
    \label{fig1}
\end{figure}

We consider a system made of  
graphene
having three regions labeled by $j = {1}, {2}, {3}$ such that the medium region is subjected to a
magnetic barrier together with a square potential  and
a mass term as geometrically
represented in Figure \ref{fig1}. Note that, the mass term
 can be owing to the sublattice symmetry breaking or can be
seen as the energy gap $\Delta = \Delta_{so}$ originating from
the spin-orbit interaction.
Near Dirac point ${\boldsymbol{K}}$, our system can be described by
the  Hamiltonian 
    \begin{equation}\lb{ham2}
        H_{j}=v_F{\boldsymbol{\sigma}}\cdot
        \left({\boldsymbol{p}}+e{\boldsymbol{A}}\right)+V_{j}(x)\mathbb{I}_2+\Delta\Theta(Lx-x^{2})\sigma_z
    \end{equation}
    where $v_{F}$ is the Fermi velocity, $
    {\boldsymbol{\sigma}} =(\sigma_{x}, \sigma_{y})$ are the  Pauli
    matrices and ${\mathbb I}_{2}$ is the $2 \times 2$ unit matrix,
     $\Theta$ is the step function. We use a
  static square potential barrier of the form
    \begin{equation}
        V_j(x)=
        \left\{%
        \begin{array}{ll}
            V_{0}, & \qquad\hbox{$0\leq x\leq L$} \\
            0, & \qquad \hbox{otherwise}. \\
        \end{array}%
        \right.
    \end{equation}
Regarding  a magnetic barrier, the relevant physics is described by a
    magnetic field translationally invariant along the $y$-direction
    $B(x, y)= B(x)$. It is associated to 
    the vector
    potential  in Landau gauge ${\boldsymbol{A}} = (0,A_{y}(x))^{T}$ since
    $\partial_{x}A_{y}(x)= B(x)$ and consequently the transverse momentum $p_{y}$ is
a conserved quantity. In our study we consider
 $\textbf{B}=B\textbf{e}_z$ along $z$-direction
     within the strip $0\leq x\leq L$ but $B=0$
    elsewhere, namely
    $   B(x,y)= B\Theta(Lx-x^{2})$,
    with  $B$ is a constant.
    Due to the continuity, we have
    the potential
    \begin{equation}
        \qquad A_{y}(x)=\left\{%
        \begin{array}{ll}
            0, & \qquad \hbox{$x<0$} \\
            Bx, & \qquad \hbox{$0\leq x\leq L$} \\
            BL, & \qquad \hbox{$x>L$}. \\
        \end{array}%
        \right.
    \end{equation}

The solutions of energy spectrum associated to \eqref{ham2} can be obtained by solving Dirac equation for the
 spinor $\Psi_j(x, y)= e^{ik_y y} (\varphi^{+}_{j}(x), \varphi^{-}_{j}(x))^T$
    \begin{equation}\label{eqva}
        \left[{\boldsymbol{\sigma}}\cdot
    \frac{1}{\hbar} \left({\boldsymbol{p}}+e{\boldsymbol{A}}\right)+\tilde{V}_{j}(x)\mathbb{I}_2+\tilde{\Delta}\Theta(Lx-x^{2})\sigma_z\right]\Psi_j(x,y)=
        \epsilon\Psi_j(x,y)
    \end{equation}
    and we have set the quantities $\epsilon=E/\hbar\upsilon_F$, $\tilde{V}=\frac{V}{\hbar v_F}$ and
    $\tilde{\Delta}=\frac{\Delta}{\hbar v_F}$.
    Consequently, in region $1$ $(x<0)$, the solution is obtained as sum of incident $ \Psi_{\sf in}(x,y) $
    and reflected $  \Psi_{\sf re}(x,y) $ waves that is
    \begin{eqnarray}\label{555}
        \Psi_1(x,y)=\frac{1}{\sqrt{2}} \begin{pmatrix}
            1\\
            e^{i\phi}
        \end{pmatrix} e^{i (k_x x+k_y y)}+\frac{r}{\sqrt{2}} \begin{pmatrix}
            1\\
            -e^{-i\phi}
        \end{pmatrix} e^{i (-k_x x+k_y y)}
    \end{eqnarray}
    where $r$ is the reflection amplitude and the
    wave vector component is defined in terms of the  incident energy $ \epsilon  $ as
    \begin{equation}\label{kxx}
    k_x=s\sqrt{\epsilon^{2}-k^{2}_{y}}.
    \end{equation}
    It is convenient for our task to express the wave vector components $ k_x $
    and $k_y$ in terms of the incident angle $\phi$.  Thus,  we can write
    \begin{equation}
        k_x=\epsilon\cos\phi, \qquad k_y=\epsilon\sin\phi
    \end{equation}

    As far as  region $3$ $(x > L)$ is concerned, we solve \eqref{eqva} to end up with  the eigenspinor $ \psi_{3}(x, y)$ for
    transmitted electron 
    \begin{equation}\label{777}
        \Psi_{3}(x,y)= 
        \frac{t}{\sqrt{2}} \begin{pmatrix}
            1\\
            e^{i\phi'}
        \end{pmatrix} e^{i k'_x x +k_y y}
    \end{equation}
    where $t$ is the transmission amplitude and we have
    \begin{equation}\label{kpx}
         k'_x=\sqrt{\epsilon^2-(k_y+\beta L)^2}
    \end{equation}
 such that
 $\beta=\frac{eB}{\hbar}$ has the inverse of length.  The transmitted angle is related to
    the wave vector components via
    \begin{equation}
        k'_x=\epsilon\cos\phi', \qquad k_y=\epsilon\sin\phi'-\beta L
    \end{equation}
    and consequently  we establish the relation
    \begin{equation}\label{111}
        \sin\phi'=\sin\phi+\frac{\beta L}{\epsilon}.
    \end{equation}

    Regarding region 2 ($0\leq x\leq L$), let us first write
    the corresponding Hamiltonian in terms of the annihilation $a^-$ and
    creation $ a^+ $ operators
    \begin{equation}\label{eq 20}
        H_{2}=\hbar v_{F}\left(%
        \begin{array}{cc}
            V^{+} & -i\sqrt{2\beta} a^{-}\\
            i\sqrt{2\beta} a^{+}  & V^{-} \\
        \end{array}%
        \right)
    \end{equation}
   such that $V^{\pm}= \tilde{V}\pm\tilde{\Delta}$ and we have
    \beq
    a^{\pm}=\frac{1}{\sqrt{2\beta}}\left(\mp\partial_{x}+k_y+x\beta\right)
    \eeq
    which satisfy the commutation relation $[a,a^\dagger]=\mathbb{I}$.
     As usual, we solve the eigenvalue equation for
$\psi_{2}(x, y)=e^{ip_{y}y}  (\varphi_{2}^+(x), \varphi_{2}^-(x))^{T}$
    \begin{equation}\label{eq 23}
        H_{2}\left(%
        \begin{array}{c}
            \varphi_{2}^+ \\
            \varphi_{2}^-\\
        \end{array}%
        \right)=\hbar v_{F}\epsilon\left(%
        \begin{array}{c}
            \varphi_{2}^+\\
            \varphi_{2}^-\\
        \end{array}%
        \right)
    \end{equation}
to obtain
    two coupled equations
    \begin{eqnarray}\label{eq 25}
        &&V^{+}\varphi_{2}^{+}-i\sqrt{2\beta}a^{-}\varphi_{2}^-=\epsilon\varphi_{2}^+\\
        &&\label{eq 26}
        i\sqrt{2\beta}a^{+}\varphi_{2}^+ +
        V^{-}\varphi_{2}^{-}=\epsilon\varphi_{2}^-.
    \end{eqnarray}
    Injecting \eqref{eq 26} into \eqref{eq 25} we end up with  a
    differential equation of second order for
    $\varphi_{2}^{+}$
    \begin{equation}
        \left(\epsilon-V^{+}\right)\left(\epsilon-V^{-}\right)\varphi_{2}^{+}=2\beta a^-
        a^{+}\varphi_{2}^{+}.
    \end{equation}
    This is in fact an equation of the harmonic oscillator and
    therefore we identify $\varphi_{2}^{+}$ with its eigenstates
    $|n-1\rangle$ associated  to the eigenvalues
    \begin{equation}\label{eq34}
        \tilde{\varepsilon}=\sqrt{2\beta n+\tilde{\Delta}^{2}}
    \end{equation}
    where we have set $\tilde{\varepsilon}=s'\left(\epsilon-\tilde{V}\right)$,
    $s'=\mbox{sign}\left(\epsilon-\tilde{V}\right)$ correspond to positive and negative
    energy solutions. From \eqref{eq 26}, we derive the second spinor component
    \begin{equation}
        \varphi^{-}_{2}=s'i\sqrt{\frac{\tilde{\varepsilon}-s'
                \tilde{\Delta} }{\tilde{\varepsilon}+s' \tilde{\Delta}}} \mid
        n\rangle.
    \end{equation}
Combing all components we can map
the eigenspinors
 in terms of
the parabolic cylinder functions $
    D_{n}(x)=2^{-\frac{n}{2}}e^{-\frac{x^{2}}{4}}H_{n}\left(\frac{x}{\sqrt{2}}\right)
$ as
    \begin{eqnarray}
        \Psi_{\sf 2}(x,y) = e^{ik_{y}y}\sum_{\pm}c^{\pm}
        \begin{pmatrix}
            \sqrt{\frac{\tilde{\varepsilon}+s'\tilde{\Delta}
                }{\tilde{\varepsilon}}}
            D_{\left(\tilde{\varepsilon}^{2}-\tilde{\Delta}^{2} \right)/2\beta-1}
            \left(\pm \sqrt{\frac{2}{\beta}}\left(\beta x+k_{y}\right)\right) \\
            \frac{ \pm i\sqrt{2\beta}}{\sqrt{\tilde{\varepsilon}
                    \left(\tilde{\varepsilon}+s' \tilde{\Delta}\right)}}
            D_{\left(\tilde{\varepsilon}^{2}-\tilde{\Delta}^{2}\right)/2\beta}
            \left(\pm \sqrt{\frac{2}{\beta}}\left(\beta x+k_{y}\right)\right) \\
        \end{pmatrix}%
       \end{eqnarray}
   with the Hermite polynomials  $ H_{n}(x) $. Note that,
    the transmission $t$ and reflection $r$ coefficients will be determined
     by using the boundary conditions  at  interfaces.

    \section{Group delay time}

    We study the group delay time in transmission and reflection
    beams around some transverse wave vector $k_y $ and incident angle  $\phi \in [0, \frac{\pi}{2}]$ fulfilling the relation \eqref{111}. This is because the wave
    incident from the right- and left-hand sides of the surface normal
    will behave differently \cite{Ghosh00}. Then, we characterize our
    waves by introducing a critical angle $\phi_{c}$ defined by
    \begin{equation}
        \phi_{c}=\arcsin\left[1-\frac{L\beta}{\epsilon}\right]
    \end{equation}
which corresponds to take $ \phi'=\frac{\pi}{2} $ in \eqref{111}. Consequently,
   an immediate conclusion takes place such that
    we have  oscillating guided modes when  $\phi< \phi_{c}$ and
    decaying or evanescent wave modes in the opposite case. As a matter of  simplicity,  we will be interested in studying  the
    case $\phi<\phi_{c}$ in the
    forthcoming analysis.

    Based on different
    considerations, we study the interesting properties of our system
    in terms of the corresponding transmission and reflection probabilities. To start we use
    the continuity of
    eigenspinors  at the interfaces $x=0$ and $x=L$ to write the boundary conditions
    \begin{equation}
            \Psi_1(0,y)=\Psi_{2}(0,y), \qquad
            \Psi_{2}(L,y)=\Psi_{3}(L,y)
    \end{equation}
which give rise to a set of equations. Then
    after a lengthy but straightforward algebra, we  show that the
    transmission  and reflection coefficients
    take the forms
    \begin{eqnarray}
        && t=\frac{2\xi_{1}\xi_{2}}{\chi} (u_{1L}v_{2L}+u_{2L}v_{1L})\cos{\phi}\\
        &&r= \frac{\Lambda}{\chi}
    \end{eqnarray}
where we have defined the quantities
    \begin{eqnarray}
        &&\chi=\left(\xi_{2}u_{20}-\xi_{1} u_{10}e^{-i\phi}\right) \left(\xi_{1}
        v_{1L}e^{i\phi'}-\xi_{2}v_{2L}\right)+\left(\xi_{1}
        u_{1L}e^{i\phi'}+\xi_{2}u_{2L}\right)\left(\xi_{2}v_{20}+\xi_{1}
        v_{10}e^{-i\phi}\right)\\
        &&\Lambda=\left(-\xi_{2}u_{20}-\xi_{1} u_{10}e^{i\phi}\right) \left(\xi_{1}
        v_{1L}e^{i\phi'}-\xi_{2}v_{2L}\right)+\left(\xi_{1}
        u_{1L}e^{i\phi'}+\xi_{2}u_{2L}\right)\left(-\xi_{2}v_{20}+\xi_{1}
        v_{10}e^{i\phi}\right)
    \end{eqnarray}
    and the following
    shorthand notation
    \begin{eqnarray}
        &&u_{1x}= D_{\left(\tilde{\varepsilon}^{2}-\tilde{\Delta}^{2}
            \right)/2\beta-1}
        \left(\sqrt{\frac{2}{\beta}}\left(\beta x+k_{y}\right)\right), \qquad
        u_{2x}=D_{\left(\tilde{\varepsilon}^{2}-\tilde{\Delta}^{2}\right)/2\beta}
        \left(-\sqrt{\frac{2}{\beta}}\left(\beta x+k_{y}\right)\right)\\
        &&v_{1x}= D_{\left(\tilde{\varepsilon}^{2}-\tilde{\Delta}^{2} \right)/2\beta-1}
        \left(\sqrt{\frac{2}{\beta}}\left(\beta x+k_{y}\right)\right), \qquad
        v_{2x}=D_{\left(\tilde{\varepsilon}^{2}-\tilde{\Delta}^{2}\right)/2\beta}
        \left(-\sqrt{\frac{2}{\beta}}\left(\beta x+k_{y}\right)\right)\\
        &&\xi_{1}=\sqrt{\frac{\tilde{\varepsilon}+s'\tilde{\Delta}
            }{\tilde{\varepsilon}}}, \qquad
        \xi_{2}=\frac{i\sqrt{2\beta}}{\sqrt{\tilde{\varepsilon}
                \left(\tilde{\varepsilon}+s' \tilde{\Delta}\right)}}.
    \end{eqnarray}
    It is easy to show the mapping
    \begin{equation}\label{pshi}
        t=|t|e^{i\varphi_{t}}, \qquad t=|r|e^{i\varphi_{r}}
    \end{equation}
    in terms
    of the phase shifts  $\varphi_{t}$ and  $\varphi_{r}$. Both relations will play a crucial rule in computing the group delay time associated to our system.

  To determine   the transmission $T$ and reflection $R$ probabilities, we introduce
    the current density, which can be found to be 
    \begin{equation}
        J=ev_F
        \psi^+\sigma_x\psi
    \end{equation}
which allows to obtain
its incident $J_{\sf in}$,  transmitted  $J_{\sf tr}$ and reflected $J_{\sf re}$
components. Then, from the relations $ T=\frac{|J_{\sf tr}|}{|J_{\sf in}|} $
and $ R=\frac{|J_{\sf re}|}{|J_{\sf in}|} $ we get
     the results
     \begin{equation}
     T=\frac{k'_x}{k_x}|t|^2, \qquad R=|r|^2.
     \end{equation}

In the next, we show  how  the above tools can be used to study   the group
    delay time in transmission and reflection. In the beginning, let us notice that a finite pulsed
    electron beam can be represented by a tempo-spatial wave packet as  weighed superposition of plane wave spinors. As a results,
    we can express
    the incident, reflected  and transmitted beam waves at
    $x=0$ 
    as  double Fourier integrals over $\omega$
    and $k_y$ \cite{Chen3x, Chen4x}. They are
    \begin{eqnarray}
        &&\label{int1}
        \Phi_{\sf in}(x,y, t)=\iint f(k_y,\omega)\ \Psi_{\sf in} (x,y) \  e^{-i\omega
            t}\ dk_yd\omega\\
   &&\label{int2}
        \Phi_{\sf re}(x,y,t)=\iint  f(k_y,\omega)\ \Psi_{\sf re} (x,y) \ e^{-i \omega
            t}\ dk_yd\omega
    \\
    &&\label{int3}
        \Phi_{\sf tr}(x,y,t)=\iint  f(k_y,\omega) \ \Psi_{\sf tr} (x,y) \ e^{-i \omega
            t}\ dk_yd\omega
    \end{eqnarray}
where the three spinors $ \Psi_{\sf in}, \Psi_{\sf re} $ and  $ \Psi_{\sf tr} $ are given in \eqref{555}
and \eqref{777}, respectively.    The frequency of wave is $\omega=E/\hbar$ and
    the angular spectral distribution is
    assumed to be  of  Gaussian shape, i.e.
    $f(k_y,\omega)=w_ye^{-w_{y}^2(k_y-\omega)^2}$ with   the
    half beam width at waist $w_y$ \cite{Beenakker}. From (\ref{int2}-\ref{int3}),
    we get the total phases
    for the reflected and transmitted waves functions at ($x=0$,
    $x=L$), which are
    \begin{equation}
    \phi_{r}=\varphi_{r}+k_yy-\omega t, \qquad \phi_{t}=\varphi_{t}+(k_y+\beta L)y-\omega t.
    \end{equation}

To  determine the Goos-H\"anchen (GH) shifts $ S_{\gamma} $
and
group delay time $ \tau_{\gamma} $, one can use
the stationary phase approximation
\cite{Steinberg1, Li1}.  In this case  $ S_{\gamma} $
is given by
    \begin{equation}\label{ghss}
        S_{\gamma}=- \frac{\partial \varphi_{\gamma}}{\partial
            k_{y}}
    \end{equation}
    where $\gamma=t, r$ stands for  transmission and reflection.
   As for $ \tau_{\gamma} $ we use
    the two derivatives of phase shifts
    with respect to the
    incident angle $\phi$ and frequency $\omega$. Then from the conditions
    \begin{equation}\label{key}
        \frac{\partial\varphi_\gamma}{\partial\phi}=0, \qquad
    \frac{  \partial\varphi_\gamma}{\partial\omega}=0
    \end{equation}
we  end up with a result
    \begin{equation}
        \tau_{\gamma}= \tau^{\varphi_{\gamma}} +
        \tau^{s_{\gamma}}
    \end{equation}
as sum of  two parts 
\begin{equation}\label{tpa}
\tau^{\varphi_{\gamma}} =   \frac{\partial \varphi_{\gamma}}{\partial
        \omega}, \qquad
\tau^{s_{\gamma}}=  \left(\frac{\partial k_y}{\partial
        \omega}\right)S_{\gamma}
\end{equation}
   such that
    $\tau^{\varphi_{\gamma}}$ represents  the time derivative of
    phase shifts, while $\tau^{s_{\gamma}}$ results from the
    contribution of  the shift $S_{\gamma}$. Since the wave
    function is involving two-component spinor then
    $\tau_{\gamma}$ can be regarded as  average of the
    group delay times of two components. Consequently,  we have in phase shifts

    \begin{eqnarray}
     \tau^{\varphi_{t}}=\hbar \frac{\partial \varphi_t}{\partial
            E}+\frac{\hbar}{2}\frac{\partial \phi'}{\partial E}, \qquad
        \tau^{\varphi_{r}}=\hbar \frac{\partial \varphi_r}{\partial E}
    \end{eqnarray}
    and in GH shifts
    \begin{equation}
        \tau^{s_{t}}=\frac{\sin\phi}{\upsilon_F}S_t, \qquad
        \tau^{s_{r}}=\frac{\sin\phi}{\upsilon_F}S_r.
   \end{equation}
These results will be numerically investigated to emphasis the basic features of our system.
     It will help us to understand the effect of
    various potential parameters  on the group delay time of a gap opening
    in graphene  magnetic barrier scattered by a square potential.

    \section{Numerical results}
   To numerically study  the  transmission probability $T$ and group delay time $\tau_{\gamma}$ for our system, we first fix some requirements needed. Indeed,  according to (\ref{kpx}) we show that
   the magnetic field $ B $ should fulfill  the  condition
\begin{equation}
 \frac{\sqrt{|(E-V)^2-\Delta^2|}}{e L \upsilon_F}<B< \frac{E}{eL \upsilon_F}
\end{equation}
in order to have a real wave vector $k'_x$. Then beyond this,
$k'_x$
will be
 imaginary, which physically entails the evanescence of the wave function inside the barrier. In contrast, when the magnetic field satisfies (\ref{kpx}), the evanescent wave function exists,  but  still propagating inside the transmission region.
 {{As a result,  for our numerical analysis, we choose  the parameters
$\phi=0$, $V=80 $ meV, $E = 125 $ meV, $\Delta=20$ meV, $L = 75 $ nm to get the
magnetic interval 0.53  T $< B < $ 1.67  T}}. 
Additionally,
    in each configuration a set of appropriate values of physical parameters will be fixed depending on the 
    plots. These will result in generating interesting information regarding our system and may shed some light on its possible technological application. In the next and  for numerical reason we use dimensionless group delay time $\tau_{t}/\tau_0$ by introducing a time scale $\tau_0= L/v_{F}$. In fact this will help to understand and determine the particle velocities when cross the barrier either greater or less than that of Fermi.

\begin{figure}[ht]
    \centering
    \subfloat[$B$=0.1 T, 0.3 T, 0.4 T]
    {
        \centering
        \includegraphics[width=8cm,height=5cm]{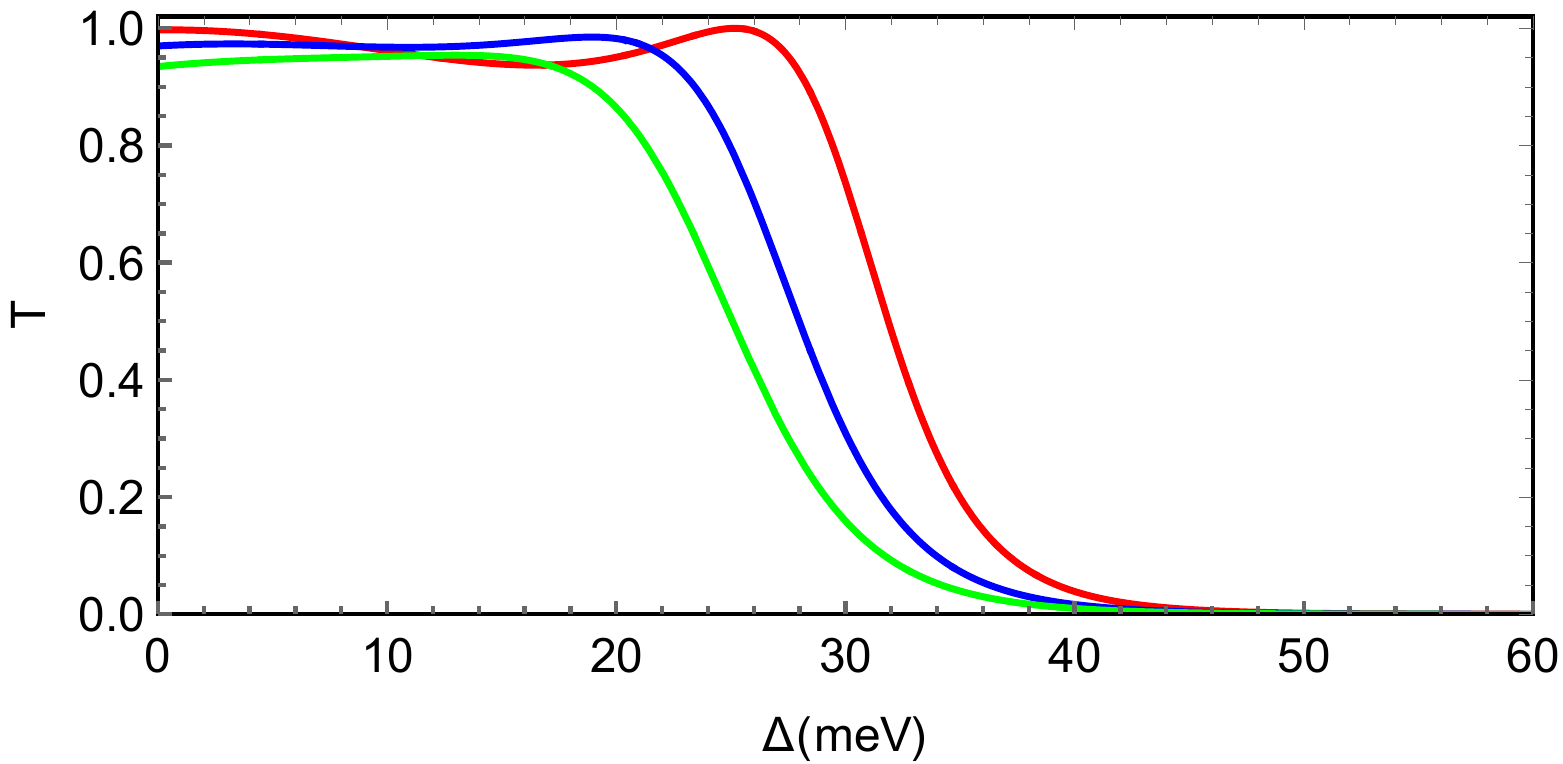}
        \label{fig6a}
    }\subfloat[$B$=0.1 T, 0.3 T, 0.4 T]{
        \centering
        \includegraphics[width=8cm,height=5cm]{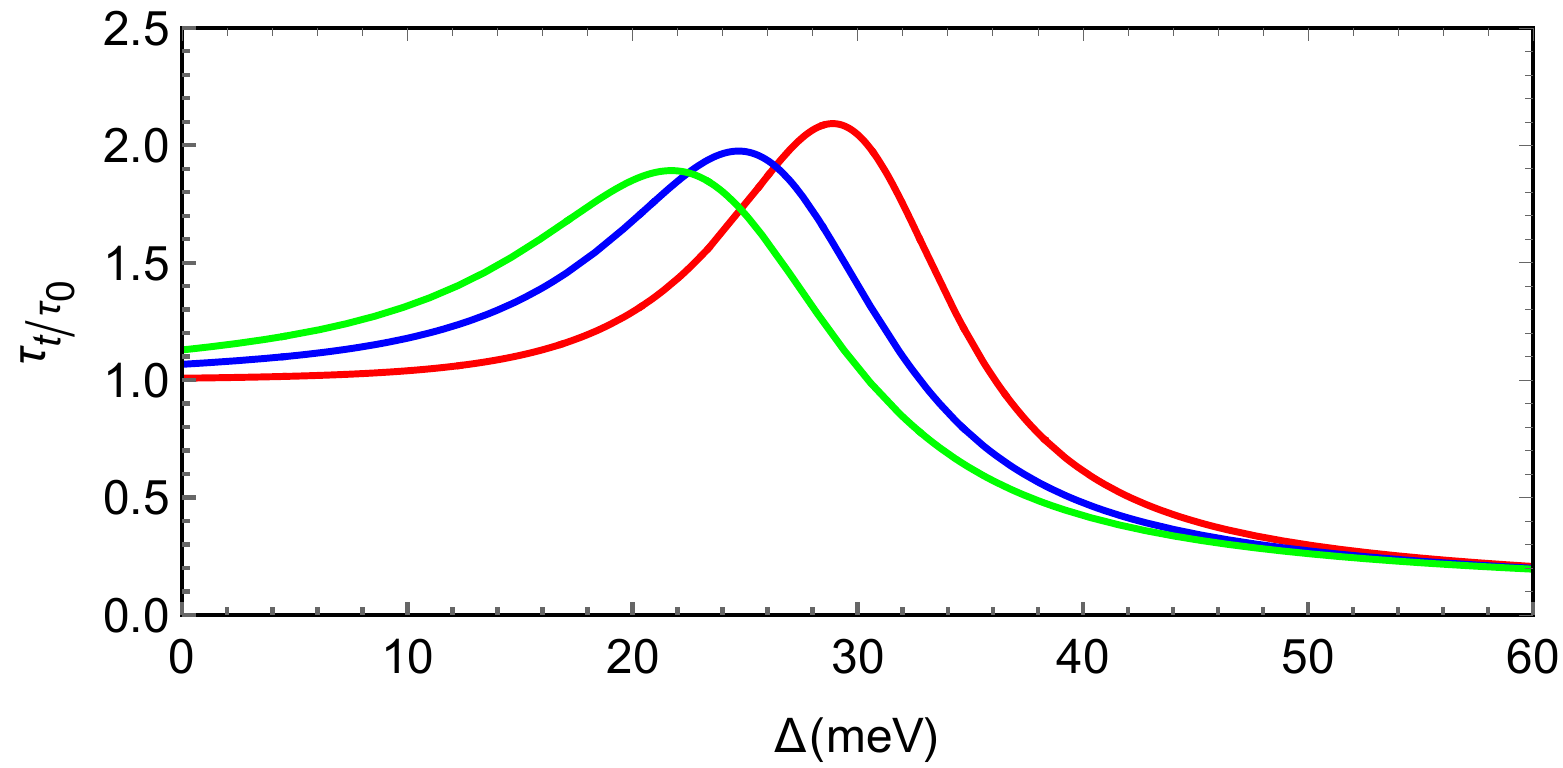}
        \label{fig6b}}
    \caption{\sf{(color online) Panel {\color{red}{(a)}}: The transmission $T$ and panel {\color{red}{(b)}}: group delay in transmission $\tau/\tau_{0}$ 
            as a function of the energy gap $\Delta$  for three values of the magnetic field
            $B=0.1$ T (red), $0.3$ T (blue), $0.4$ T (green)  with
            $\phi=0$, $E=120$ meV, $V=80$ meV, $L=70$ nm. }}

    \label{fig6}
\end{figure}

    Figure \ref{fig6} shows
the transmission probability $T$ and group delay time $\tau_{t}/\tau_0$ as a function of the
    energy gap $\Delta$ for  three values of magnetic field
      {{$B=0.1$ T (red), $0.3$ T (blue), $0.4$ T (green)}} with
    barrier height $V_{0}=80$ meV and width $L=70$ nm.
In panel \ref{fig6a},
we distinguish two energy zones such that
in the first one $\Delta$ $<E-V_{0}$, there is full transmission (Klein tunneling) even with the increase of
$\Delta$ because the wave vector inside the barrier is real corresponding to the
transmission mode. Whereas  in the second zone
$\Delta$ $>E-V_{0}$, we notice that
$T$ decays
exponentially as long as  $\Delta$  increases and later on it
approaches to  zero because the wave vector $k_x$ inside the barrier becomes imaginary giving rise to
an evanescent mode. Additionally, according the values taken by $B$ we observe that
$T$
decreases rapidly as $B$ increased.
        In panel \ref{fig6b},
   {{we observe that for $\Delta$ $=0$ and $B=0.1$ T (red)
    the particles propagate through the barrier with the Fermi velocity $v_{F}$ ($\tau_{t}/\tau_{0}=1$)}} but by increasing
    $\Delta$ one sees that $\tau_{t}/\tau_{0}$ start to increase slowly to reach a maximum and after that it decreases rapidly toward a constant value, which becomes independent of  $\Delta$.
    We notice that $ B $ affects the  behavior of $\tau_{t}/\tau_{0}$ because it decreases when $ B $ increases.

\begin{figure}[ht]
    \centering
    \subfloat[$\phi=25, 30, 35$]{
        \centering
        \includegraphics[width=8cm,height=5cm]{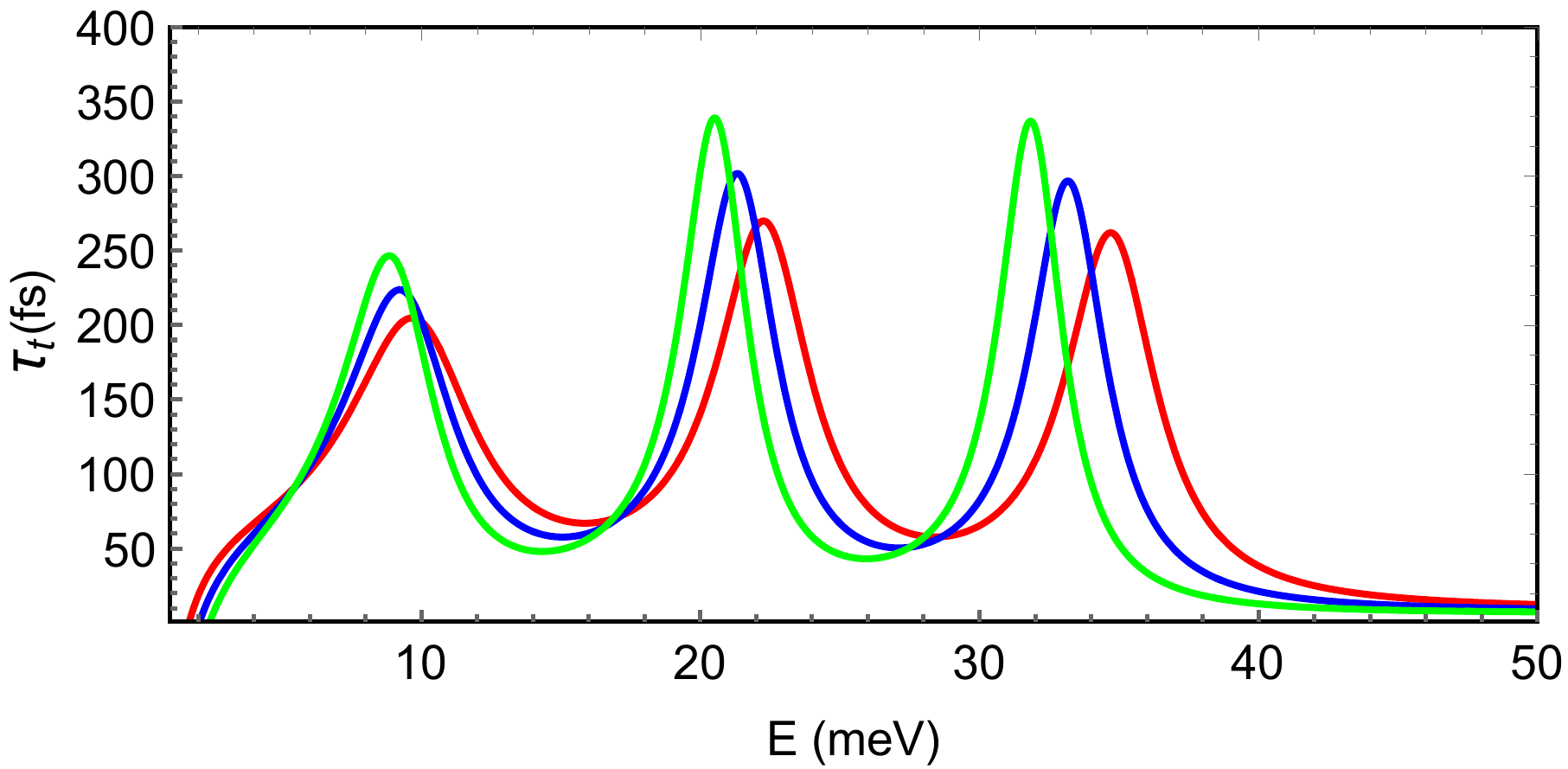}
        \label{fig2a}
    }\subfloat[$L=65\ \mbox{nm}, 85 \ \mbox{nm}, 122 \ \mbox{nm}$]{
        \centering
        \includegraphics[width=8cm,height=5cm]{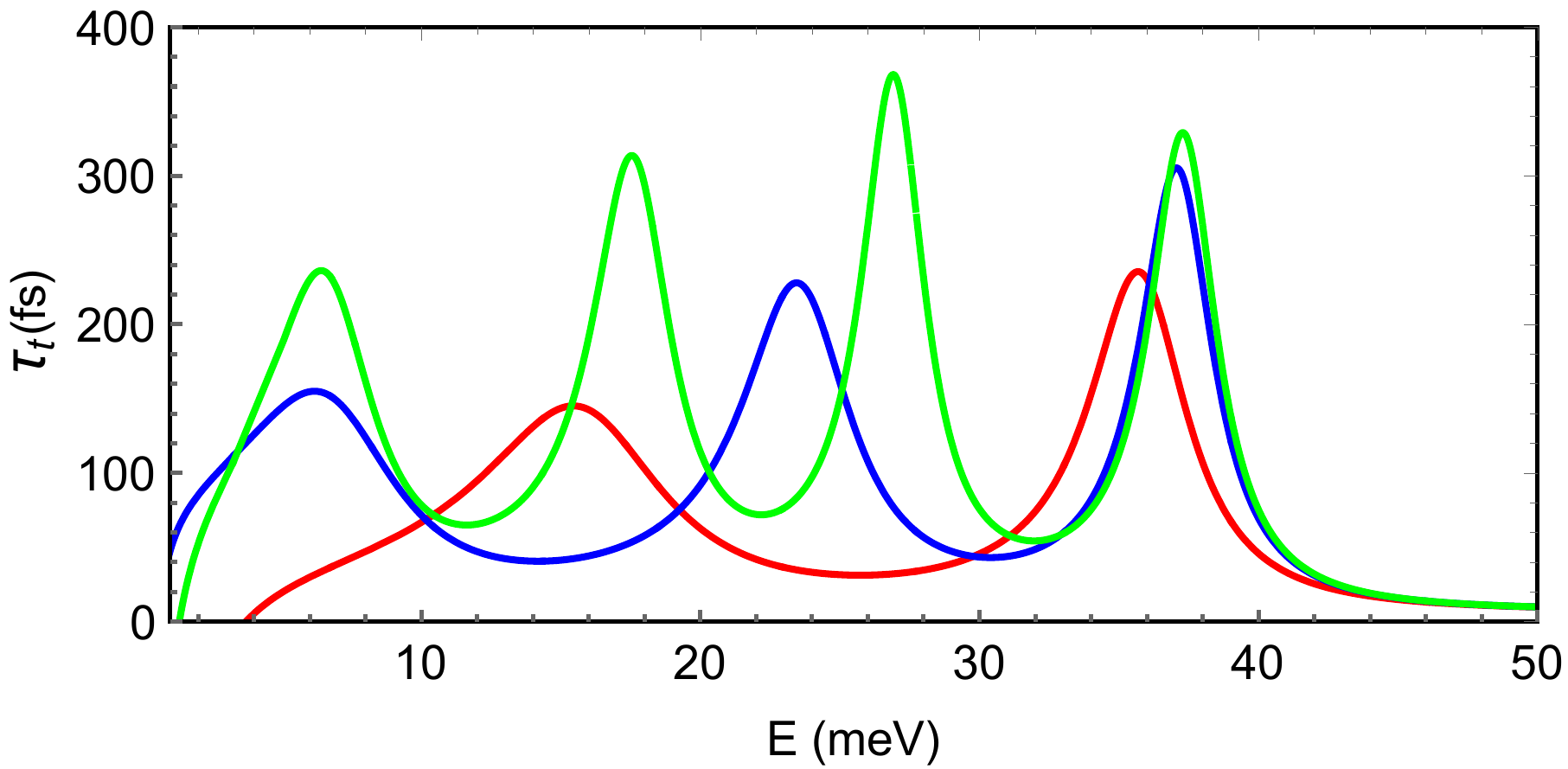}
        \label{fig2b}}\\
    \subfloat[$B=0.4\ \mbox{T}, 0.5\ \mbox{T}, 0.6\ \mbox{T}$]{
        \centering
        \includegraphics[width=8cm,height=5cm]{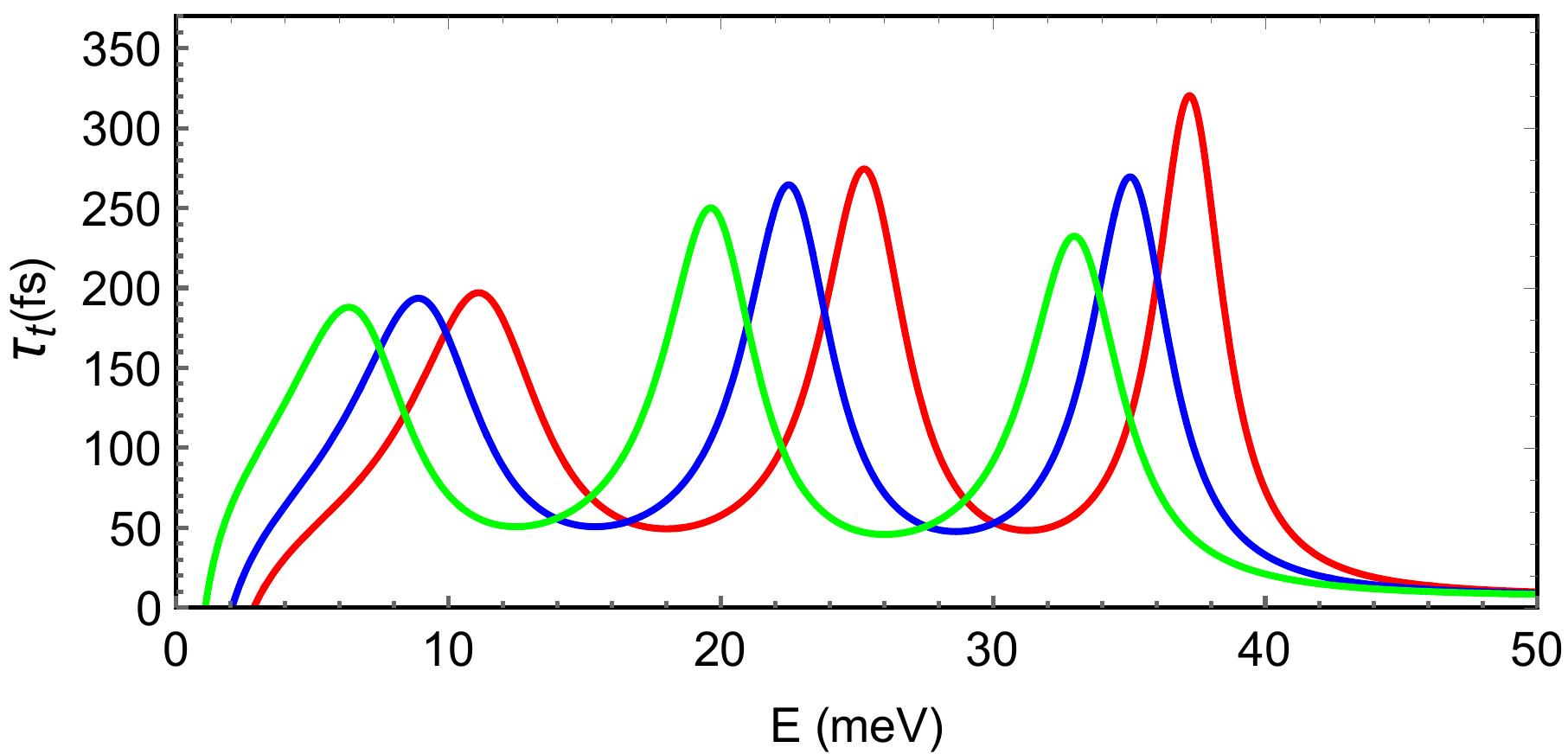}
        \label{fig2c}
    }\subfloat[$\Delta=10\ \mbox{meV}, 20\ \mbox{meV}, 30 \ \mbox{meV}$]{
        \centering
        \includegraphics[width=8cm,height=5cm]{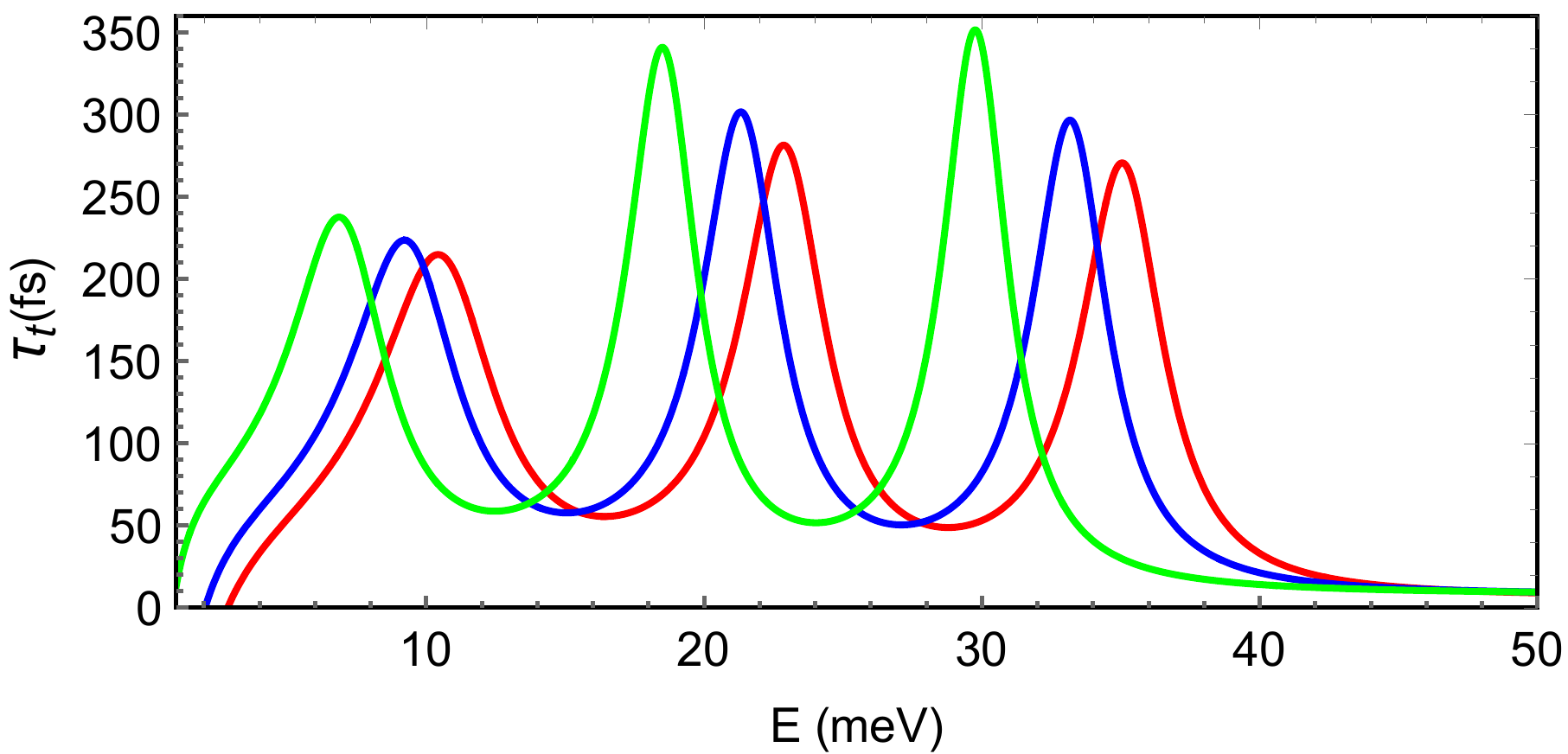}
        \label{fig2d}}
    \caption{\sf{(color online) The group delay time in transmission  $\tau_{t}$ as
            a function of the incident energy $E$ for the barrier height $V_{0}=80$ meV. Panel
            {\color{red}{(a)}}: $\phi$= 25 (red), 30 (blue), 35 (green)  with $B=0.5$~T, $L=100$ nm,
            $\Delta=20$ meV.
        Panel   {\color{red}{(b)}}: $L$=65 nm (red), 85 nm (blue), 122
            nm (green) with $B=0.4$ T, $\Delta=10$ meV, $\phi=30$.
        Panel   {\color{red}{(c)}}: $B$=0.4 T (red), 0.5 T (blue), 0.6 T (green) with
            $L=95$ nm, $\Delta=10$ meV, $\phi=30$.
        Panel   {\color{red}{(d)}}: $\Delta$= 10 meV (red), 20 meV (blue), 30
            meV (green) with $B=0.5$ T, $L=100$ nm, $\phi=30$.}}
    \label{fig2}
\end{figure}

    In Figure \ref{fig2} we plot the group delay time $\tau_{t}$ as a
    function of the incident energy $E$  for various physics parameters.
    Indeed, panel \ref{fig2a} illustrates the case of
     three values of the incident angle $\phi=25$ (red), $30$ (blue), $35$ (green) for
$V_{0}=80$ meV,  {{$B=0.5$ T, $L=100$ nm and
$\Delta=20$ meV}}. We observe that
      $\tau_{t}$ oscillates with the
    increase of  $E$ and also its peak
  increases as long as
     $\phi$ increases, but when $E$ becomes larger
    than a critical value $E_{c}=40$ meV,
    $\tau_{t}$  approaches quickly zero.
{Panel \ref{fig2b}
    shows the influences of
    the barrier widths $L=65$ nm (red), $95$ nm (blue), $122$ nm (green) with
    $V_{0}=80$ meV, $B=0.4$ T, $\phi=30$ and $\Delta=10$ meV.
    When  $E<E_{c}$, $\tau_{t}$  shows oscillatory behavior with different amplitudes
    and the number of oscillations increases with the increase  of $L$. Another important remark is that when $L$ becomes large enough  $\tau_{t}$ becomes independent of $  E $. We notice that when  $E$ is larger than $E_{c}$, $\tau_{t}$ goes to stabilize  at zero whatever the value of $L$.
    In panel \ref{fig2c} we consider
    three values of  magnetic field  $B=0.4$ T
    (red), $0.5$ T (blue), $0.6$ T (green) for $V_{0}=80$ meV, $L=95$ nm, $\phi=30$ and $\Delta=10$ meV. It is clearly seen that $B$
    affects the oscillatory behavior of $\tau_{t}$
    because its  peaks get decreased by increasing $ B $ and their positions shifted
    clockwise as well.
    Panel \ref{fig2d}  presents three values of the energy gap  $\Delta=10$ meV (red),
    $20$ meV (blue), $\Delta=30$ meV (green),
    for $V_{0}=80$ meV,  {{$L=100$ nm, $\phi=30$ and $ B=0.5 $ T}}.
 As for the case $E<E_{c}$, we observe  that
    $\tau_{t}$ oscillates with different amplitudes.  Now  by increasing
     $\Delta$, we notice that there is an important increase of
    $\tau_{t}$ and  their peaks
    also get affected in position and value.}

\begin{figure}[ht]
    \centering
    \subfloat[$\phi=25$, $30$, $35$]{
        \centering
        \includegraphics[width=8cm,height=5cm]{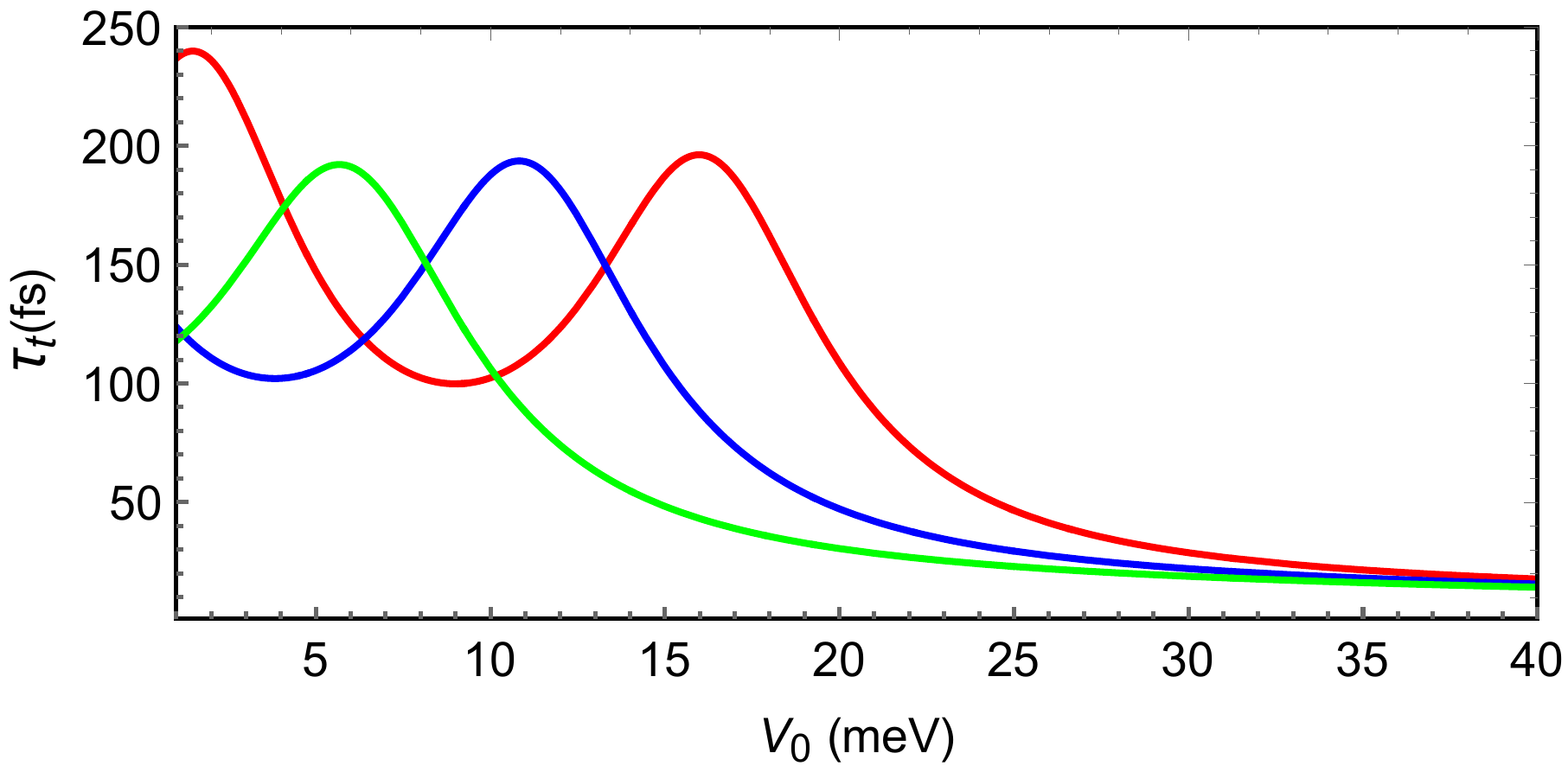}
        \label{fig3a}
    }\subfloat[$L=80$ nm, 90 nm, 95 nm]{
        \centering
        \includegraphics[width=8cm,height=5cm]{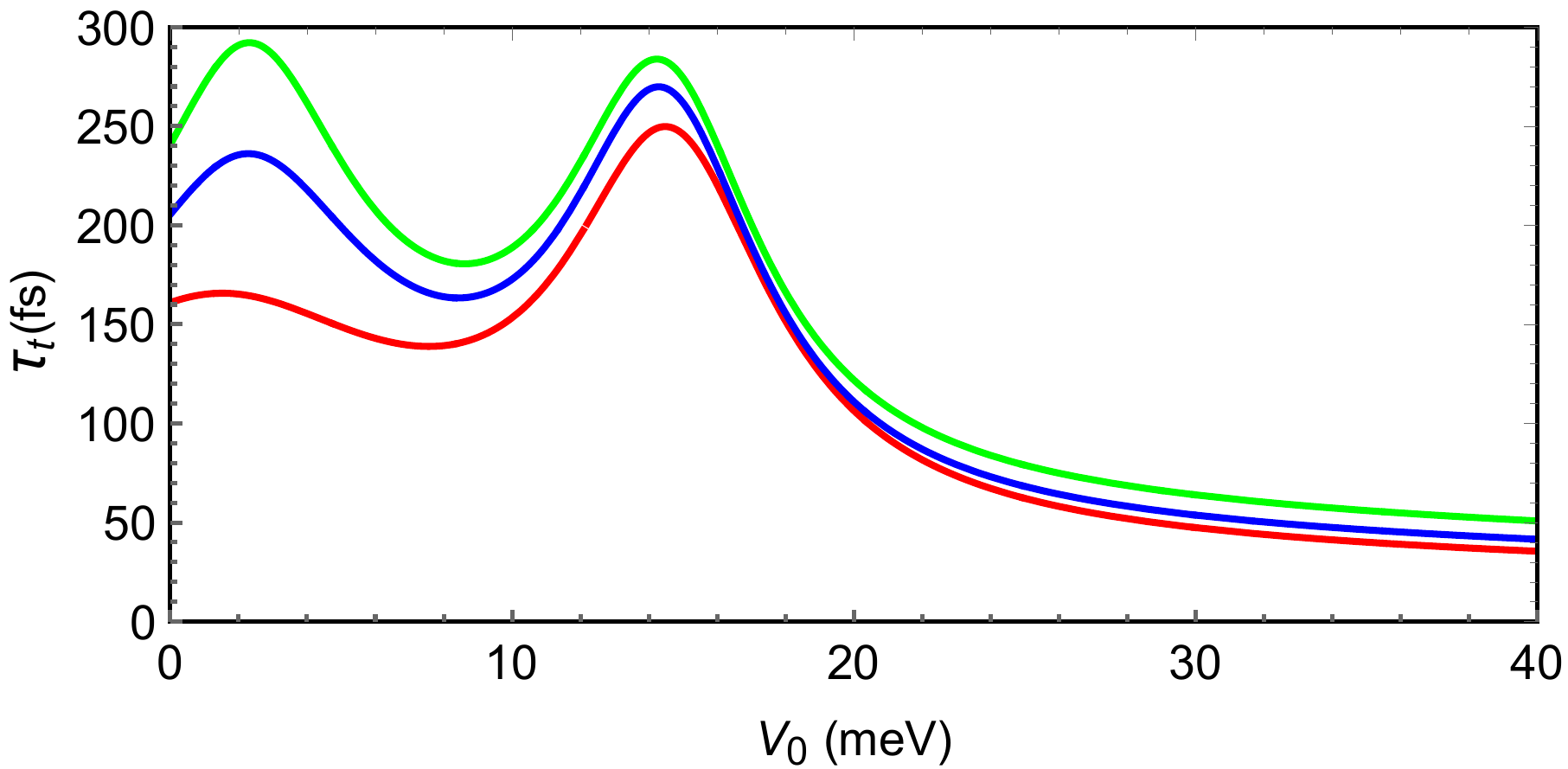}
        \label{fig3b}}\\
    \subfloat[$B=0.4$ T, $0.5$ T, $0.6$ T]{
        \centering
        \includegraphics[width=8cm,height=5cm]{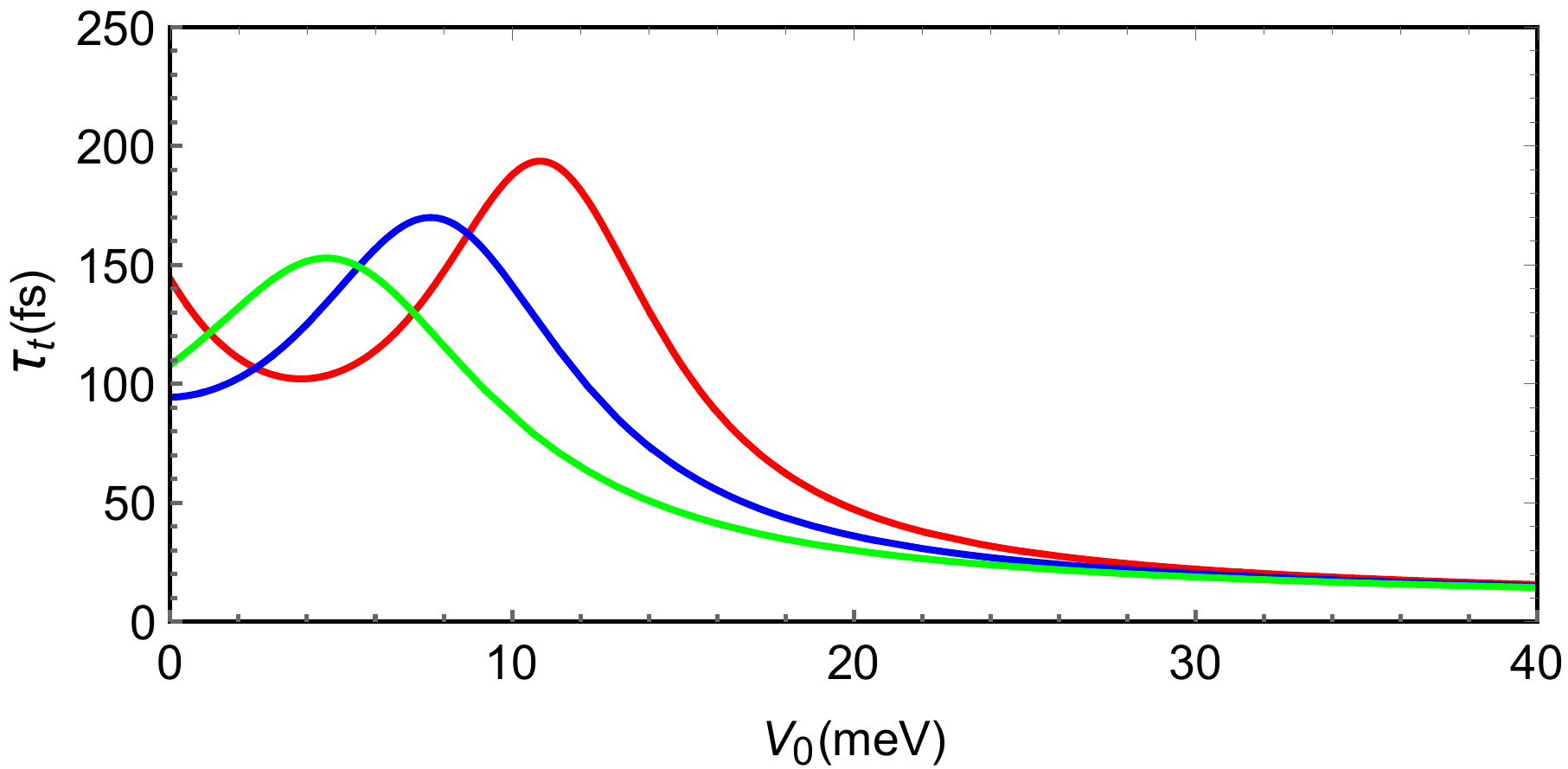}
        \label{fig3c}
    }\subfloat[$\Delta=10$ meV, $20$ meV, $30$ meV]{
        \centering
        \includegraphics[width=8cm,height=5cm]{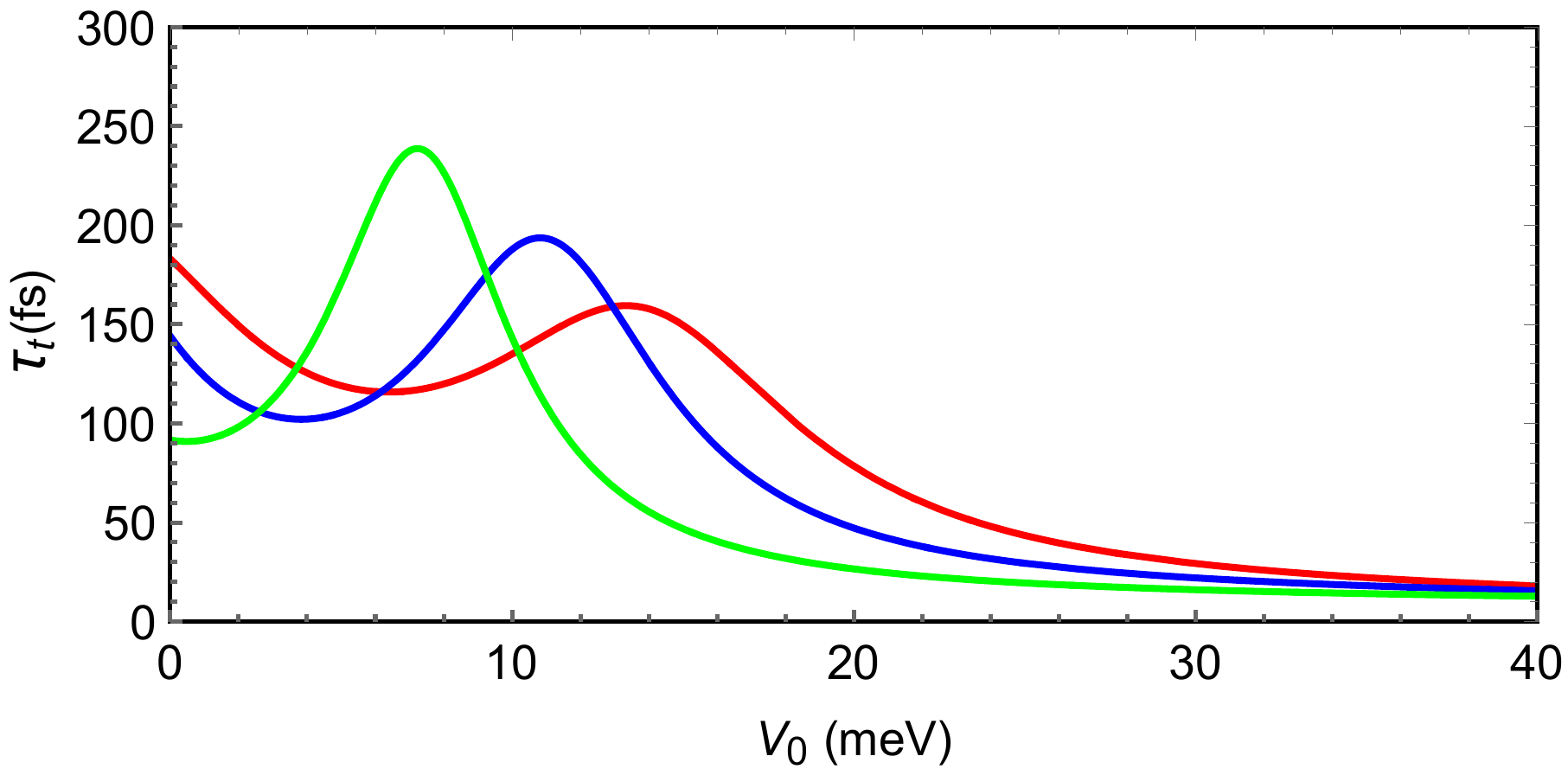}
        \label{fig3d}}
    \caption{\sf{(color online) The group delay time in transmission as a
            function of barrier height $V_{0}$ for the incident energy $E=80$ meV.
            Panel {\color{red}{(a)}}:  $\phi$= 25 (red), 30 (blue), 35 (green), $B=0.5$ T, $L=100$ nm,
            $\Delta=20$ meV. Panel {\color{red}{(b)}}:  $L$=(80 nm,
            $90$ nm, $95$ nm), $\phi=30$, $B=0.4$ T, $\Delta=20$
            meV. Panel {\color{red}{(c)}}: $B=0.4$ T (red), $0.5$ T (blue), $0.6$
            T (green), $L=100$ nm, $\Delta=20$. Panel {\color{red}{(d)}}:
            $\Delta$=10 meV (red), $20$ meV  (blue), $30$ meV (green), $\phi=30$, $B=0.5$ T, $L=100$
            nm.}}
    \label{fig3}
\end{figure}

Figure \ref{fig3} presents the group delay time  $\tau_{t}$ as function of the  barrier height $V_{0}$ for  $E=80$ meV and under
 suitable choices of physical parameters. 
Panel \ref{fig3a} present the effect of three values of the
incident angle $\phi=25$ (red), $30$ (blue), $35$ (green) for
 {{$B=0.5$ T}}, $L=100$ nm and $\Delta=20$ meV. We observe that
$\tau_{t}$ oscillates by decreasing when  $V_{0}$ increases and
goes to a constant value after some critical value of $V_{0}$.
Also its peaks decrease as long as  $\phi$ increases because for $
V_{0}=0 $, $\tau_{t}$ has maximal value for $\phi=25$ compared to
$\phi=30$ and $\phi=35$. It is interesting to stress  that for
different values of $\phi$, $\tau_{t}$  does not oscillate in the
same manner. Moreover, for  $V_{0} \geq 40$ meV one sees that
$\tau_{t}$ becomes not sensitive to any increase of $V_{0}$ and
converges to  a constant value. In panel \ref{fig3b} we take three
values of the barrier width $L=80$ nm (red), $90$ nm (blue), $95$
nm (green) for $\phi=30$, $B=0.4$ T and $\Delta=20$ meV. It is
clear that
  $\tau_{t}$ represents an oscillatory behavior and its peaks
oscillate in the same manner but their values increase with  $L$. Furthermore as for $V_{0} \geq
40$ meV,
   $\tau_{t}$
     becomes independent of  $L$ and shows
 saturation to a  constant value.
     Panel \ref{fig3c} shows
     the effect of
     three values of
    magnetic field
    $B=0.4$ T (red), $0.5$ T (blue), $0.6$ T
    (green) for
    $L=100$ nm, $\phi=30$
    and $\Delta=20$ meV. We notice that 
    $\tau_{t}$
    oscillates with the increase of  $V_{0}$ and also
     is modulated by the presence of $B$ because  when $ B $ increases $\tau_{t}$ decreases. We observe that when $V_{0}$ exceeds  $25$ meV,  $\tau_{t}$ is
    not influenced and remains constant whatever the value taken by
    $V_{0}$ and  $B$.
    In  panel \ref{fig3d}  we consider the influence of
    three values of
     energy gap
    $\Delta=10$ meV (red), $20$ meV (blue),
    $30$ meV (green)
    for $\phi=30$, {{$B=0.5$}} T, $L=100$
    nm. It shows that   the peak of
     $\tau_{t}$ increases by increasing $\Delta$ and $V_{0}$. One can see that when  $V_{0}\geq 30$ the group delay time
    $\tau_{t}$ saturates to a constant value. We conclude that $\Delta$ moderates the variation of  $\tau_{t}$ toward a tunable way.

\begin{figure}[ht]
    \centering
    \subfloat[]{
        \centering
        \includegraphics[width=8cm,height=5cm]{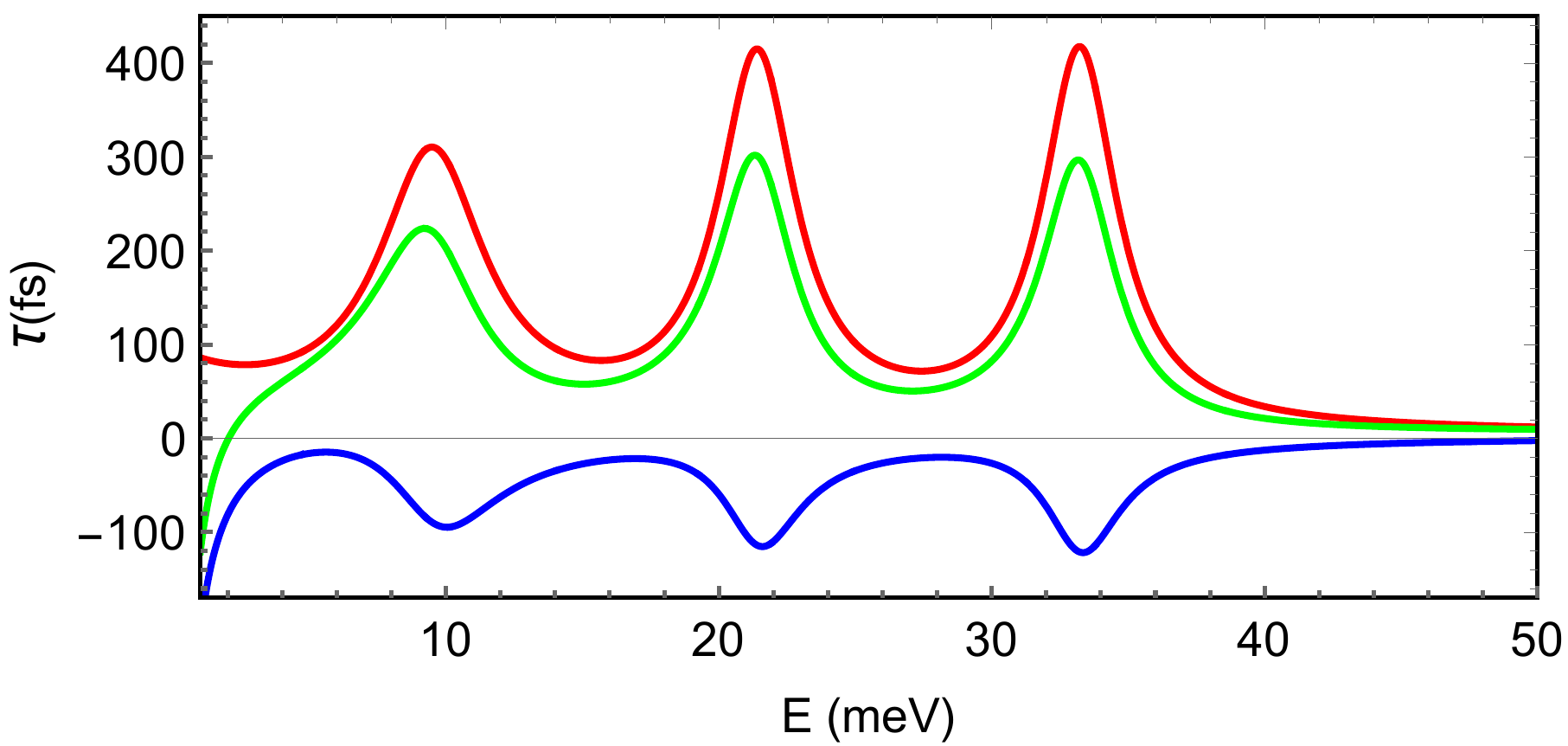}
        \label{fig4a}
    }\subfloat[]{
        \centering
        \includegraphics[width=8cm,height=5cm]{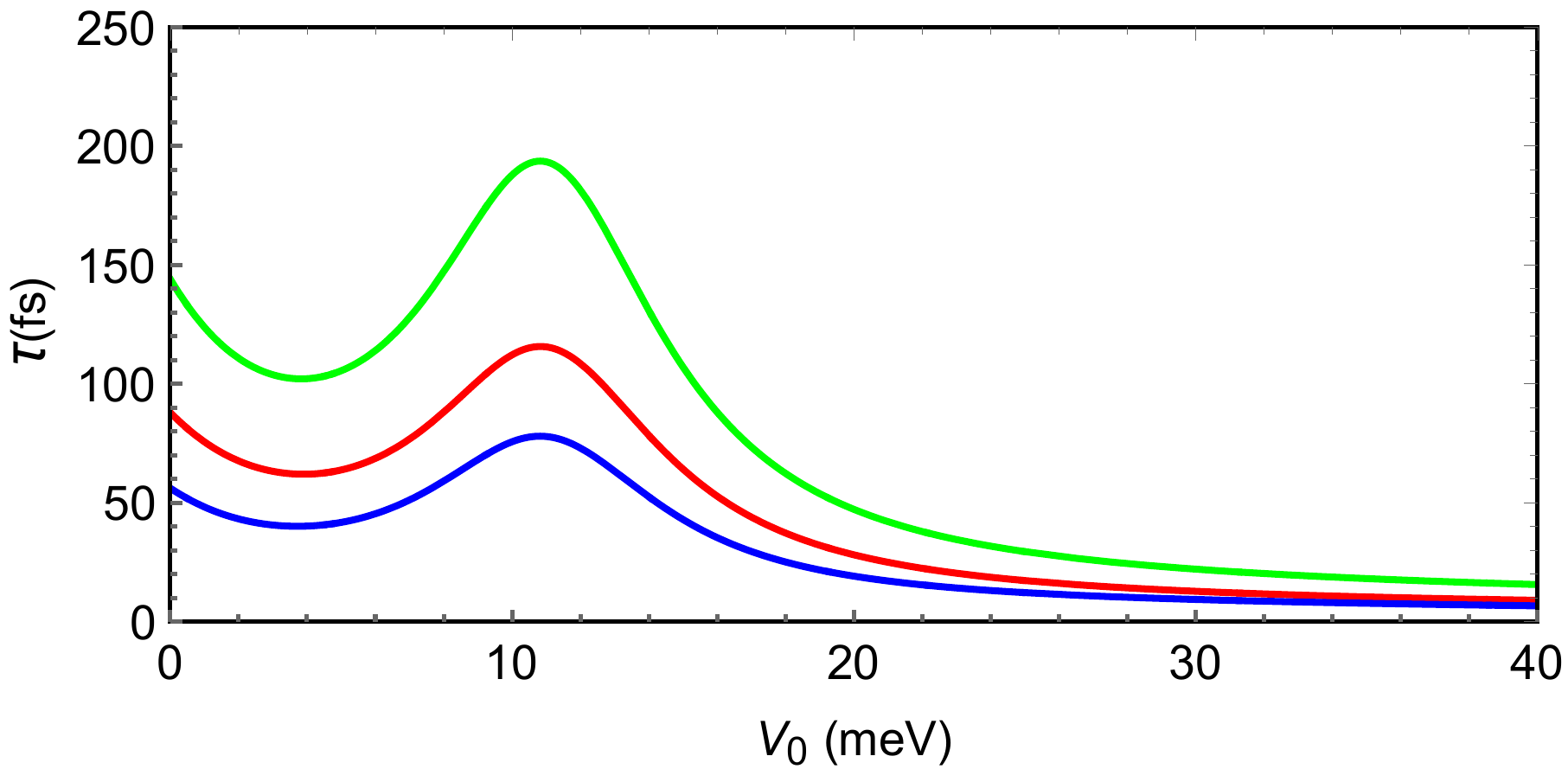}
        \label{fig4b}}
    \caption{\sf{(color online) The group delay time $\tau_{r}= \tau^{s_{r}} + \tau^{\varphi_{r}}$
            in
            reflection as a function of the incident energy $E$ and the  potential barrier $V_{0}$. Panel {\color{red}{(a)}}: $V_{0}=80$ meV
            and panel {\color{red}{(b)}}: $E=80$
            meV for $\phi=30$, $B=0.5$ T, $L=100$ nm, $\Delta=20$ meV,  with
            $\tau_{r}$ (green), $\tau^{s_{r}}$ (blue), $\tau^{\varphi_{r}}$ (red).}}
    \label{fig4}
\end{figure}

Figure \ref{fig4} presents the group delay time in reflection $\tau_{r}$
as a function of the incident energy $E$ and
barrier height $V_{0}$  with the configuration $L=100$ nm, $\phi=30$,   $\Delta=20$.
We recall that $\tau_{r}$
contains two parts such that  the phase group
delay $\tau^{\varphi_{r}}$ and that resulted from GH shifts $\tau^{s_{r}}$.
In
panel \ref{fig4a} we choose  $V_{0}=80$ meV and observe that
$\tau^{s_{r}}$ (red) and $\tau^{\varphi_{r}}$ (blue) are oscillating differently because
$\tau^{s_{r}}$ shows positive behavior and $\tau^{\varphi_{r}}$ negative one. The resulting
$\tau_{r}$   (green) oscillates from negative for small range of $ E $
  to positive for  a large range. Now in
panel \ref{fig4b} we choose $E=80$  meV
and observe that all contribution to $\tau^{\varphi_{r}}$ are oscillating positively
by showing only one peak and decrease quickly to zero as long as $ V_0 $ increase.

\begin{figure}[ht]
    \centering
    \subfloat[$\Delta=0$ meV, 9 meV, 13 meV]{
        \centering
        \includegraphics[width=8cm,height=5cm]{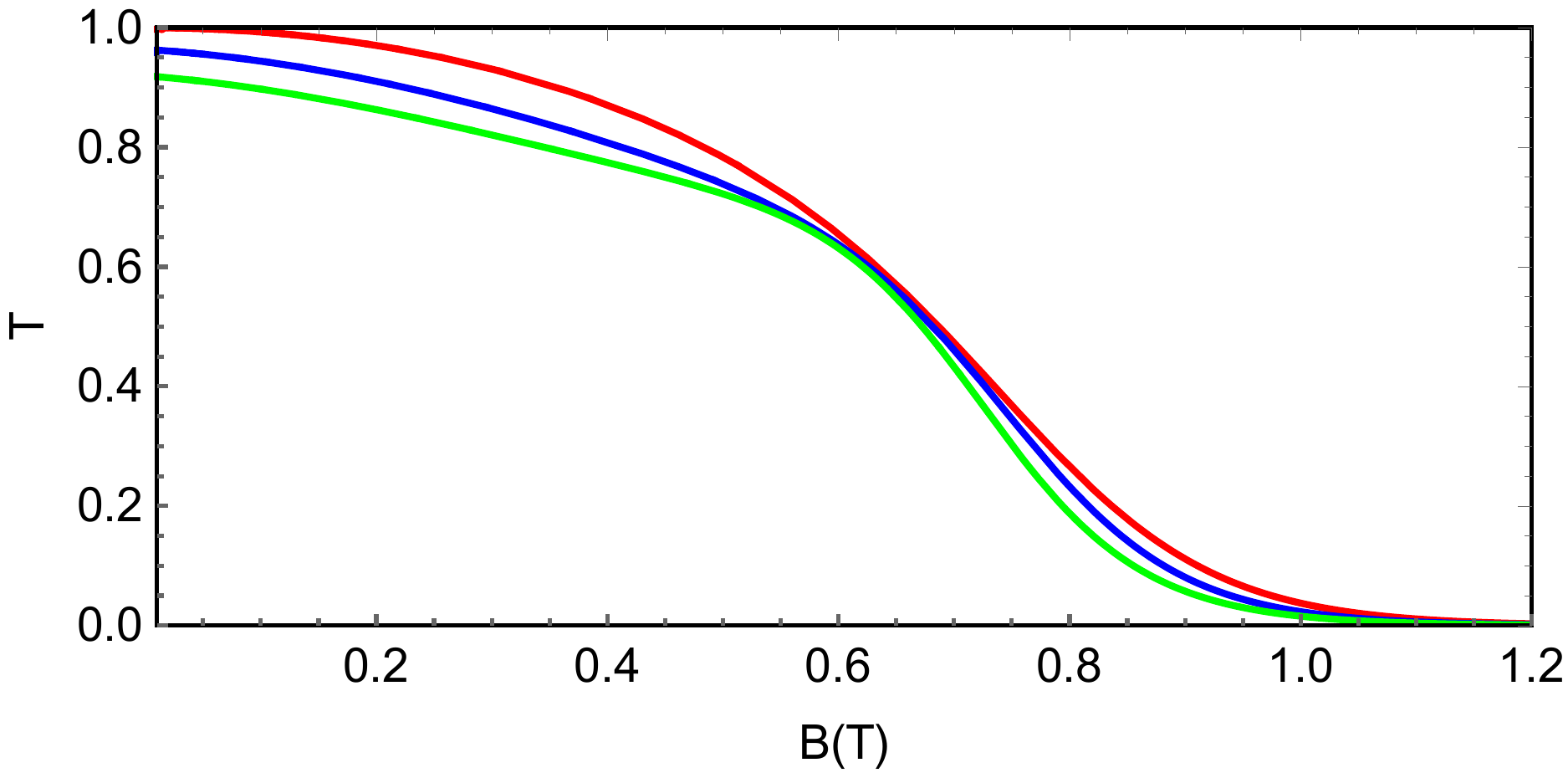}
        \label{fig5a}
    }\subfloat[$\Delta=0$ meV, 9 meV, 13 meV]{
        \centering
        \includegraphics[width=8cm,height=5cm]{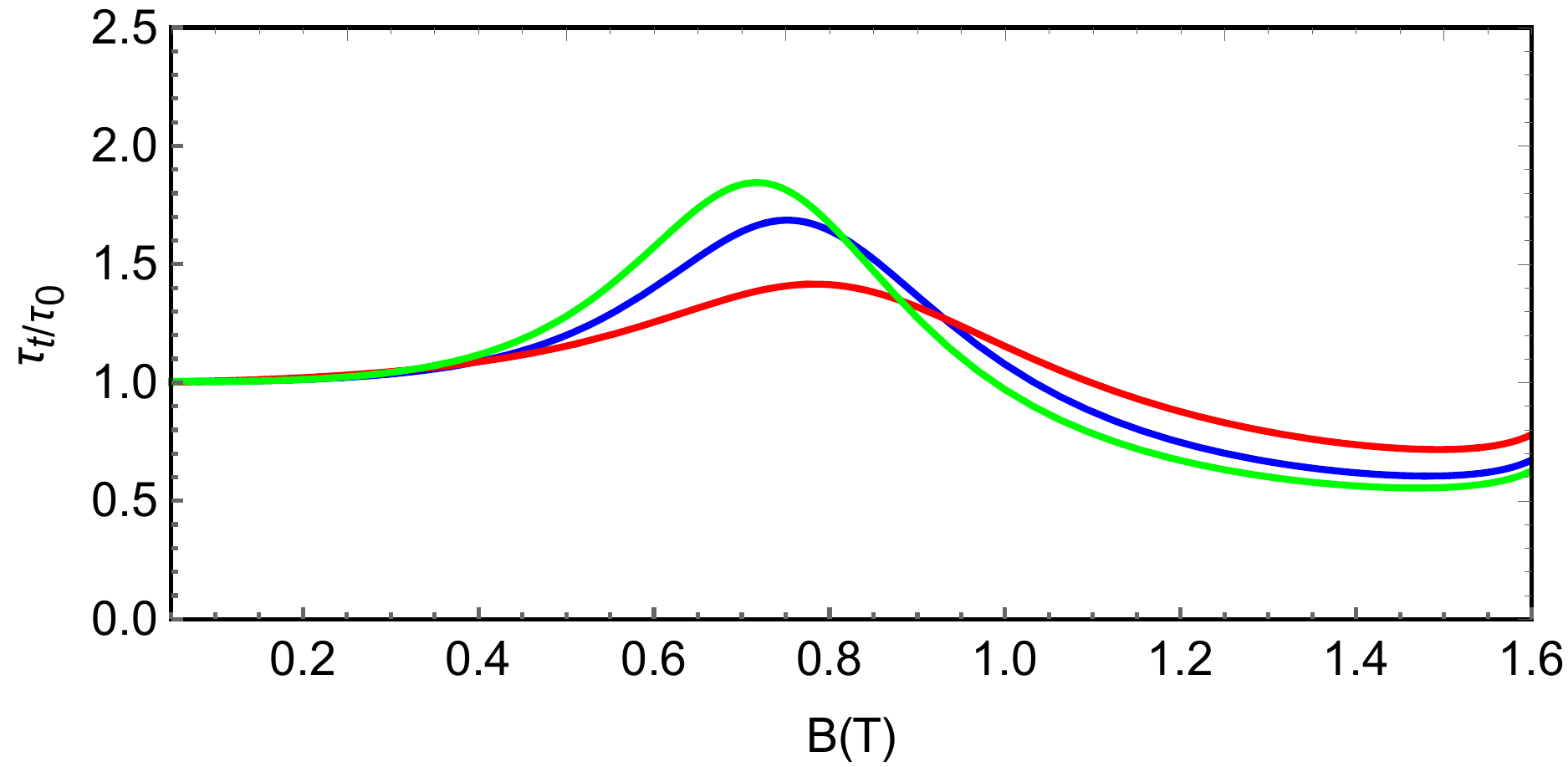}
        \label{fig5b}}
    \caption{\sf{(color online)
            Panel   {\color{red}{(a)}}:  transmission $T$ and panel {\color{red}{(b)}}: group delay time in transmission $\tau/\tau_{0}$
            as a function of the magnetic field $B$ for three values of energy gap
            $ \Delta $=0 meV (red), 9 meV (blue), 13 meV (green)  with
            $\phi=0$, $E=125$ meV, $V=80$ meV, $L$=75 nm. }}
    \label{fig5}
\end{figure}

    In Figure \ref{fig5} we plot the transmission $T$ and
    group delay time  $\tau_{t}/\tau_{0}$
     as a function of
    the magnetic field $B$ for three values of  energy gap  $\Delta=0$
    meV (red), $9$ meV (blue), $13$
    meV (green) with $\phi=0$, $E=125$
    meV, $V=80$ meV and $L=75$ nm. For null $\Delta$ (red)
    panel
    \ref{fig5a} tells us that
     $T$ decays exponentially toward zero as long as $B$ increases.
     Now for no-null $\Delta$ (blue, green) we observe that  $ T $ shows the same behavior as before except that  $\Delta$ acts by diminishing the amplitude of
     $ T $, which means that
     %
    $ T $ decreases rapidly with the increase of $\Delta$.
 In panel \ref{fig5b}, {{we observe that for the barrier static $B$ $=0$
    the particles propagate through the barrier with the Fermi velocity $v_{F}$ ($\tau_{t}/\tau_{0}=1$)}}, we notice that  for the magnetic field less than $0.4$ T, the particles propagate through the barrier with the Fermi velocity $v_{F}$,
     but when $B$ is in the range [$0.4 $ T, $1 $ T],   $\tau_{t}/\tau_{0}$ is subluminal, which
       can be changed from subluminal to superluminal. With the increase of $B$, we observe that
     $\tau_{t}/\tau_{0}$   decreases and becomes less than $L/v_{F}$, i.e. $\tau_{t}<\tau_{0}$. Note that
       for different values of  $\Delta$, the peak  of $\tau_{t}/\tau_{0}$   increases as long as  $ B $ increases.
     As we increase the energy gap  $\Delta$ we observe that the
     oscillations set in much earlier.

    \section{Conclusion}

    We have studied
    the tunneling time in   graphene
    magnetic barrier scattered by  a scalar potential and subject to a mass term.
    Taking into account the advantage of an opening gap in the energy  spectrum we have analyzed
    the corresponding group delay time. To do that we have first
    solved  Dirac equation
    to obtain the eigenspinors and used the boundary conditions at interfaces together with current density to determine the transmission and reflection probabilities. 
    After establishing a link between  our eigenspinors and beam waves, we have showed that the group delay time has two contributions resulted from phase shifts  and that of Goos-Hänchen (GH).
    More precisely,
     we have derived analytic expressions for the group
    delay time by taking into account the lateral displacement resulting
    from the angular spread of the incident electron wave packet. In fact, a
    total group $\tau_{t}$  involving  two parts such as the phase
    group delay $\tau^{\varphi_{t}}$ and the group delay contributing
    from the lateral GH shifts $\tau^{s_{t}}$ is obtained.


Subsequently, we have numerically analyzed the group delay time
by considering various choice of the physical parameters.
Indeed,  to start we have fixed the magnetic interval that allowed us to
generate interesting results.
Particularly in the propagating case, we have showed that the
group delay time is greatly enhanced by transmission resonances
thanks to the presence of energy gap. Additionally, at different
occasion we have noticed that  at some large values of the
involved physical parameters  the group delay time becomes
insensible of the increase of them and converges toward constant
values. As a result,
 the group delay time in transmission
$\tau_{t}/\tau_{0}$  can be
controlled by tuning on  $\Delta$ in orienting experiments to engineer new systems
for potential applications.


    \section*{Acknowledgment}
    The generous  support provided by the Saudi Center for Theoretical Physics (SCTP) is highly appreciated by all authors.

\end{document}